\documentclass[%
reprint,
superscriptaddress,
pra,
floatfix,
]{revtex4-2}

\usepackage{graphicx}
\usepackage{siunitx}
\usepackage{xcolor}
\usepackage{physics}

\usepackage{hyperref}
\hypersetup{pdfstartview={FitH},pdfpagemode={UseNone},
            colorlinks,linkcolor=blue, citecolor=blue, urlcolor=blue,
            bookmarksopen=true, pdfnewwindow=true}
\usepackage[all]{hypcap}

\begin{document}


\title{Directionally Tunable Co- and Counter-Propagating Photon Pairs \newline from a Nonlinear Metasurface}

\author{Maximilian~A.~Weissflog}
\email{maximilian.weissflog@uni-jena.de}
\affiliation{Institute of Applied Physics, Abbe Center of Photonics, Friedrich Schiller University Jena, Albert-Einstein-Straße 15, Jena, 07745, Germany}
\affiliation{Max Planck School of Photonics, Hans-Knöll-Straße 1, Jena, 07745, Germany}

\author{Jinyong~Ma}
\author{Jihua~Zhang}
\author{Tongmiao~Fan}
\affiliation{ARC Centre of Excellence for Transformative Meta-Optical Systems (TMOS), Department of Electronic Materials Engineering, Research School of Physics, The Australian National University, Canberra, ACT 2600, Australia}

\author{Thomas~Pertsch}
\affiliation{Institute of Applied Physics, Abbe Center of Photonics, Friedrich Schiller University Jena, Albert-Einstein-Straße 15, Jena, 07745, Germany}
\affiliation{Max Planck School of Photonics, Hans-Knöll-Straße 1, Jena, 07745, Germany}
\affiliation{Fraunhofer Institute for Applied Optics and Precision Engineering IOF, Albert-Einstein-Straße 7, Jena, 07745, Germany}

\author{Dragomir~N.~Neshev}
\affiliation{ARC Centre of Excellence for Transformative Meta-Optical Systems (TMOS), Department of Electronic Materials Engineering, Research School of Physics, The Australian National University, Canberra, ACT 2600, Australia}

\author{Sina~Saravi}
\affiliation{Institute of Applied Physics, Abbe Center of Photonics, Friedrich Schiller University Jena, Albert-Einstein-Straße 15, Jena, 07745, Germany}

\author{Frank~Setzpfandt}
\affiliation{Institute of Applied Physics, Abbe Center of Photonics, Friedrich Schiller University Jena, Albert-Einstein-Straße 15, Jena, 07745, Germany}
\affiliation{Fraunhofer Institute for Applied Optics and Precision Engineering IOF, Albert-Einstein-Straße 7, Jena, 07745, Germany}

\author{Andrey~A.~Sukhorukov}
\email{andrey.sukhorukov@anu.edu.au}
\affiliation{ARC Centre of Excellence for Transformative Meta-Optical Systems (TMOS), Department of Electronic Materials Engineering, Research School of Physics, The Australian National University, Canberra, ACT 2600, Australia}

\begin{abstract}
    Nonlinear metasurfaces have recently been established as a new platform for generating photon pairs via spontaneous parametric down-conversion. While for classical harmonic generation in metasurfaces a high level of control over all degrees of freedom of light has been reached, this capability is yet to be developed for photon pair generation. In this work, we theoretically and experimentally demonstrate for the first time precise control of the emission angle of photon pairs generated from a nonlinear metasurface. Our measurements show angularly tunable pair-generation with high coincidence-to-accidental ratio for both co- and counter-propagating emission. The underlying principle is the transverse phase-matching of guided-mode resonances with strong angular dispersion in a nonlinear lithium niobate metagrating. We provide a straightforward design strategy for photon pair generation in such a device and find very good agreement between the calculations and experimental results. Here we use all-optical emission angle tuning by means of the pump wavelength, however the principle could be extended to modulation via the electro-optic effect in lithium niobate. In sum, this work provides an important addition to the toolset of sub-wavelength thickness photon pair sources.
\end{abstract}


\maketitle

\section{Introduction} 

Nonlinear nanostructures and metasurfaces enable the generation of light with tailored properties in an ultra-compact, flat form-factor \cite{krasnokNonlinearMetasurfacesParadigm2018}. In the classical regime, control over various degrees of freedom such as the generated frequency \cite{liuAlldielectricMetasurfaceBroadband2018}, polarization state \cite{weissflogFarFieldPolarizationEngineering2022}, optical wavefront \cite{wangNonlinearWavefrontControl2018,grinblatNonlinearDielectricNanoantennas2021}, and orbital angular momentum \cite{liNonlinearMetasurfaceSimultaneous2017} have been realized for various harmonic generation processes in nonlinear nanostructures. With the emergence of quantum-entangled photons as key-enabler for applications like secure communication protocols \cite{loSecureQuantumKey2014} or quantum-enhanced imaging and sensing \cite{gilabertebassetPerspectivesApplicationsQuantum2019}, leveraging the ability of metasurfaces to control nonlinearly generated light has great potential \cite{Poddubny:2020-147:QuantumNonlinear, solntsevMetasurfacesQuantumPhotonics2021, Wang:2022-38:PT, Neshev:2023-26:NPHOT, Kan:2023-2202759:ADOM}. Indeed, recently the generation of photon pairs using spontaneous parametric down-conversion (SPDC) has been observed in nonlinear metasurfaces \cite{santiago-cruzPhotonPairsResonant2021,santiago-cruzResonantMetasurfacesGenerating2022,sonPhotonPairsBidirectionally2023,zhangSpatiallyEntangledPhoton2022} as well as isolated nanoparticles \cite{marinoSpontaneousPhotonpairGeneration2019a,duongSpontaneousParametricDownconversion2022a,saerensBackgroundFreeNearInfraredBiphoton2023}. First demonstrations included the control of the spectrum of down-converted photons. The usually broad spectrum of photon pairs from non-phasematched, thin nonlinear crystals \cite{okothMicroscaleGenerationEntangled2019, santiago-cruzEntangledPhotonsSubwavelength2021} can be reshaped and controlled by introducing resonances via nanostructuring \cite{santiago-cruzEntangledPhotonsSubwavelength2021,santiago-cruzResonantMetasurfacesGenerating2022}. Furthermore, thin-crystal based SPDC sources allow the generation of various polarization entangled quantum states \cite{weissflogTunableTransitionMetal2023} and also the tuning of the degree of polarization entanglement \cite{sultanovFlatopticsGenerationBroadband2022a}, a concept that can be extended when leveraging resonances in nanoresonators \cite{weissflogNonlinearNanoresonatorsBell2024}.

\begin{figure}[ht!]
\centering
\includegraphics[width=1\columnwidth]{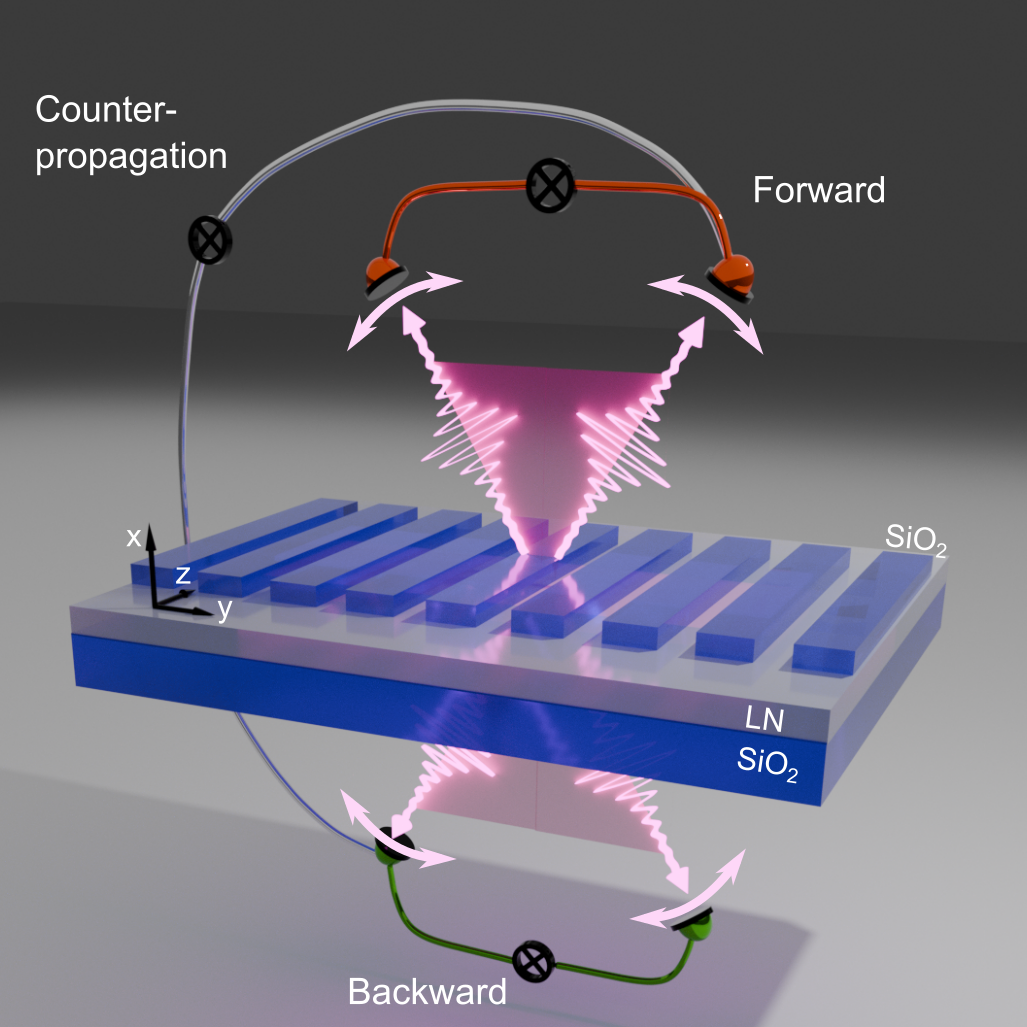}
\caption{Principle of a metagrating that generates angularly tunable photon pairs in co- and counter-propagating geometry.}
\label{fig0}
\end{figure}

However, many capabilities of nonlinear nanostructures demonstrated in the classical regime are yet to be developed for quantum processes in metasurfaces. An example is the control of the spatial emission direction: for classical frequency up-conversion, various concepts for flexible tuning of the free-space emission direction exist \cite{difrancescantonioAllopticalFreespaceRouting2023}. In contrast, while nonlinear metasurfaces based on Mie-type or quasi-BIC resonances can generate photon pairs in different spatial configurations like backward \cite{santiago-cruzPhotonPairsResonant2021}, forward \cite{santiago-cruzResonantMetasurfacesGenerating2022} and counter-propagating \cite{sonPhotonPairsBidirectionally2023} emission, potentially with high directionality \cite{Nikolaeva:2021-43703:PRA}, no ability to actively control the spatial properties was demonstrated experimentally. In this work, we close this gap by theoretically and experimentally demonstrating a nonlinear lithium niobate metagrating that offers precise control over the free-space emission direction of photon pairs, see the conceptual sketch in Fig.~\ref{fig0}. Our nonlinear metagrating leverages transverse phase-matching of guided-mode resonances with strong angular dispersion \cite{wangTheoryApplicationsGuidedmode1993a} for efficient pair-generation \cite{zhangSpatiallyEntangledPhoton2022,zhangPhotonPairGeneration2023,maPolarizationEngineeringEntangled2023}.
    
We further show that this tuning mechanism works both for co- and counter-propagating photon pairs. Importantly, on the conceptual side, we present in this work a straightforward, largely analytical design strategy for a nonlinear metasurface that generates spatially tunable co- and counter-propagating photon pairs.
\newpage
The experimental results are in very good agreement with our analytic predictions. On the practical side, our photon pair source reaches a coincidence-to-accidental ratio of 7500 for counter-propagating pair-generation, a very high value for the class of (sub-) wavelength thickness sources \cite{santiago-cruzPhotonPairsResonant2021,santiago-cruzResonantMetasurfacesGenerating2022,sonPhotonPairsBidirectionally2023,guoUltrathinQuantumLight2023,weissflogTunableTransitionMetal2023,saerensBackgroundFreeNearInfraredBiphoton2023,okothMicroscaleGenerationEntangled2019,duongSpontaneousParametricDownconversion2022a}. This is an important step towards more practically usable nanoscale pair-sources, which generally suffer from a high noise level due to photoluminescence and therefore low state purity \cite{sultanovTemporallyDistilledHighDimensional2023}.

\section{Results \& Discussions}

\subsection{Metasurface Design for Directionally Tunable Photon Pair Generation}
    
\begin{figure*}[ht!]
\centering
\includegraphics[width=0.8\textwidth]{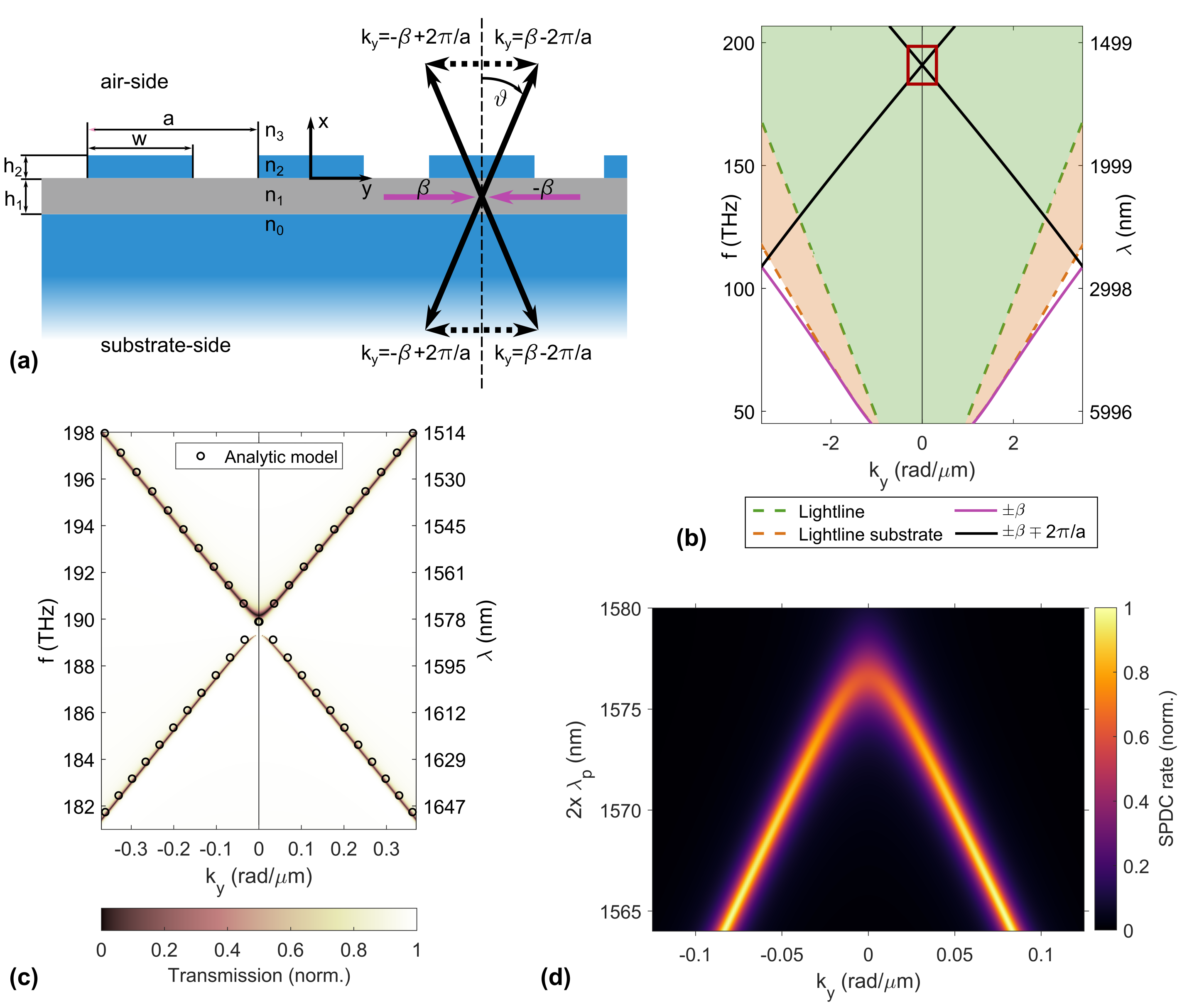}
\caption{(a) Schematic of the geometry of the employed metagrating consisting of a lithium niobate thin-film (grey) on a silicon dioxide (SiO$_2$, blue) substrate and covered by a SiO$_2$ grating. The geometrical parameters are $h_1$=308~nm, $h_2$=210~nm, $w$=560~nm and $a$=892~nm. The metagrating supports counter-propagating modes with propagation constant $\pm|\beta|$, which can couple to free-space modes with transverse wavevector $k_y$ under a propagation angle $\vartheta$ via the additional momentum provided by the grating. (b) Band structure of TE-modes in the meta-grating. The $\pm1^{st}$ orders (black solid lines) are above the light-line for both the air- and substrate-side (green and orange dashed lines and shaded areas, respectively) and therefore allow photon emission into both half-spaces. (c) Zoom-in to the region of the band-gap at $k_y$=0. The analytically calculated band-dispersion, assuming a vanishing index modulation (black circles), is overlaid with the numerically computed transmission spectrum using rigorous coupled wave analysis (RCWA) for the finite index modulation. (d) Numerical simulation of the evolution of photon pair transverse wavenumber that defines the emission angle for varying signal and idler wavelength $\lambda_s=\lambda_i=2\times\lambda_p$.
\label{fig1}}
\end{figure*}

Generally, in photon pair generation via SPDC, a pump photon with frequency $\omega_p$ splits into signal and idler photons with frequencies $\omega_{s,i}$, where the energy is conserved such that $\hbar\omega_p=\hbar\omega_s+\hbar\omega_i$. In this work, we consider a metasurface with sub-wavelength thickness, where the longitudinal phase-matching condition is relaxed \cite{okothMicroscaleGenerationEntangled2019}. The transverse pump momentum in the transversely extended structure will be conserved by the signal and idler momenta, i.e. $k_p^{\perp}=k_s^{\perp}+k_i^{\perp}$, where $k_n^{\perp}=(k_{y,n}, k_{z,n})$ and $n=p,s,i$ indicates the pump, signal or idler photon.

The main objective of this work is two-fold: the design of a nonlinear metasurface that can both generate photon pairs highly tunable in their emission angle and emit them into a co-propagating and also counter-propagating configuration. For this, a nonlinear device that shows high angular dispersion and allows photon pair emission into the same or two opposing half-spaces (air-side or substrate-side) is needed.

We will first analytically demonstrate that a nonlinear metasurface consisting of a grating structure coupled to a nonlinear slab-waveguide perfectly combines both these conditions.
We use a similar base design as in \cite{zhangSpatiallyEntangledPhoton2022}, a silicon dioxide (SiO$_2$) grating with period $a$, duty cycle $DC=w/a$ and height $h_2$ on top of an $x$-cut nonlinear lithium niobate (LN) waveguide with height $h_1$ (see Fig.~\ref{fig1}(a)). The cladding is air, therefore $n_3=1$. The linear transmission through such a device exhibits sharp resonances, so-called guided-mode resonances (GMR) \cite{wangTheoryApplicationsGuidedmode1993a, Overvig:2022-2100633:LPR}, which occur whenever an impinging lightwave is coupled to a waveguide mode. Accordingly, if a pair of signal and idler photons is generated via SPDC in the nonlinear waveguide layer, we can leverage such GMRs to couple them out into a free-space mode propagating under a specific angle.

Let us first consider a weak grating, where the index modulation $\Delta n=n_\mathrm{2}-n_\mathrm{3}\rightarrow0$ vanishes. Although this is not strictly the case for a SiO$_2$ grating with air cladding, this approximation allows us to analytically predict the emission angle of the photon pairs by finding a solution for the angular dispersion of the GMR resonances \cite{wangTheoryApplicationsGuidedmode1993a, Overvig:2022-2100633:LPR}. In the weak grating limit, the propagation constant $\beta$ of a guided mode can be calculated analytically by approximating the grating region as a homogeneous layer with averaged refractive index $n_\mathrm{avg}=DC\times n_2+(1-DC)\times n_3$ \cite{gambinoINVITEDReviewDielectric2022}. Note that in our $x$-cut LN film, the largest LN nonlinear tensor component $\chi^{(2)}_{zzz}$ mediates down-conversion of a $z$-polarized pump photon into a pair of $z$-polarized signal and idler photons. This corresponds to TE polarization in the multilayer waveguide with $j=1,...,N$ layers. The characteristic equation for TE-modes is \cite{leeAnalyticalDeterminationMultilayer2007}

\begin{equation}
    1=-\frac{q_N}{q_{N+1}}\tanh(q_N h_N+\psi_N),
    \label{eq:eigenvalue_equation}
\end{equation}
which is derived from the transverse boundary condition between the $N$-th layer and the cladding. It contains the transverse wave-vector $q_j=\sqrt{\beta^2-(2\pi/\lambda)^2 n^2_j}$, the free-space wavelength $\lambda$, the refractive index $n_j$, and the phase $\tanh(\psi_{j+1})=q_j/q_{j+1}\tanh(q_j h_j+\psi_j)$ for all $j=1,...,N$ layers. Based on the boundary condition for the initial phase $\tanh(\psi_{1})=q_0/q_{1}$ the solution of Eq.~\eqref{eq:eigenvalue_equation} is straightforward. Refer to section S1 of the supplementary material for more details. For any mode with propagation constant $\beta$ that solves Eq.~\eqref{eq:eigenvalue_equation}, a counter-propagating mode with propagation constant $-\beta$ exists, see the purple arrows in Fig.~\ref{fig1}(a) and the purple graphs in Fig.~\ref{fig1}(b). A photon propagating in the mode $\pm\beta$ can be coupled out of the waveguide if its transverse momentum is above the light line of the respective half-space.
\newpage
Since we are aiming for a source of counter-propagating photons along $\pm y$ in free-space, out-coupling both to the air-side and the substrate-side needs to be possible. The light lines for free space and the substrate are marked with green and orange dashed lines, respectively, in Fig.~\ref{fig1}(b). We choose now a set of parameters for waveguide and grating constant, such that the bands corresponding to the first grating orders $\beta-2\pi/a$ and $-\beta+2\pi/a$ cross $k_y=0$ close to the telecom wavelength range at $\approx\SI{1570}{nm}$. The analytically calculated band structure for TE modes assuming a vanishing index modulation $\Delta n\rightarrow0$ is shown in Fig.~\ref{fig1}(b). 

For finding the accurate size of the bandgap at the mode crossing point $k_y=0$ for a finite index modulation, we use rigorous coupled wave analysis (RCWA) \cite{hugoninRETICOLOSoftwareGrating2023} to simulate the transmission spectrum for incident plane waves, see the zoomed-in Fig.~\ref{fig1}(c). The very good agreement between the weak grating approximation and the rigorous calculation underpins that the analytical model is indeed an excellent guide for analyzing the angular dispersion of the metagrating. Note that at $k_y=0$ the transmission line at the branch below the band-gap vanishes (long wavelength side). This is because the corresponding eigenmode at $k_y=0$ has an antisymmetric electric field profile that does not couple to an incident plane wave. In this work, all measurements will be performed in the branch above the bandgap (short wavelength side).

The dispersion relations in Figs.~\ref{fig1}(b),(c) further demonstrate that out-coupling to both sides and therefore counter-propagating pair-emission is possible. The transverse wave vector $k_{s,i}^{\perp}$ of the signal and idler photon directly depends on the propagation constant $\beta_{s,i}$ of the guided modes and thus on the signal/idler frequencies $\omega_{s,i}$. We use a normally incident pump beam with zero transverse momentum $k^{\perp}_{p}=0$, which means the signal and idler transverse momenta have to satisfy $k_{y,s}=-k_{y,i}$. Given the mode dispersion diagram in Fig.~\ref{fig1}(b), this condition, as well as energy conservation, is satisfied for two frequency degenerate, counter-propagating modes $\pm\beta$. Therefore, we can select the emission angle of the photon pair all-optically by tuning the pump frequency $\omega_p=2\times\omega_{s,i}$, controlling in which two counter-propagating modes the signal and idler photons will be generated.

After these linear design considerations, we proceed to a calculation of the nonlinear down-conversion process. For highly multimodal, nanoscale devices, modeling the quantum process of a spontaneous decay into signal and idler photon is most efficiently treated using a Green's function approach \cite{Poddubny:2016-123901:PRL, marinoSpontaneousPhotonpairGeneration2019a, weissflogNonlinearNanoresonatorsBell2024}. However, for a metagrating supporting only one mode in the frequency region of interest and the further restriction to frequency degenerate SPDC with $\omega_s=\omega_i$, the quantum-classical correspondence between sum-frequency generation and SPDC \cite{heltHowDoesIt2012,lenziniDirectCharacterizationNonlinear2018,Poddubny:2020-147:QuantumNonlinear} can be efficiently used together with coupled-mode theory, to model the down-conversion process \cite{zhangSpatiallyEntangledPhoton2022}. The basic idea of quantum-classical correspondence is to infer the probability of down-conversion from a certain pump mode characterized by $\{\omega_p,\vec{k}_p\}$ into two signal/idler modes $\{\omega_{s,i},\vec{k}_{s,i}\}$, by calculating the efficiency of classical frequency up-conversion from $\{\omega_{s,i},-\vec{k}_{s,i}\}$ to $\{\omega_p,-\vec{k}_p\}$. As a validation of our metagrating design for angularly tunable SPDC, we calculate the down-conversion probability via the correspondence principle for varying signal/idler emission angles towards the air-side with pump incidence from the substrate-side considering frequencies $\omega_p=2\times\omega_{s,i}$, see Fig.~\ref{fig1}(d). We confirm that the emission angles of the photon pairs closely follow the dispersion of the GMRs.
    
\begin{figure*}[ht!]
\centering
\includegraphics[width=0.75\textwidth]{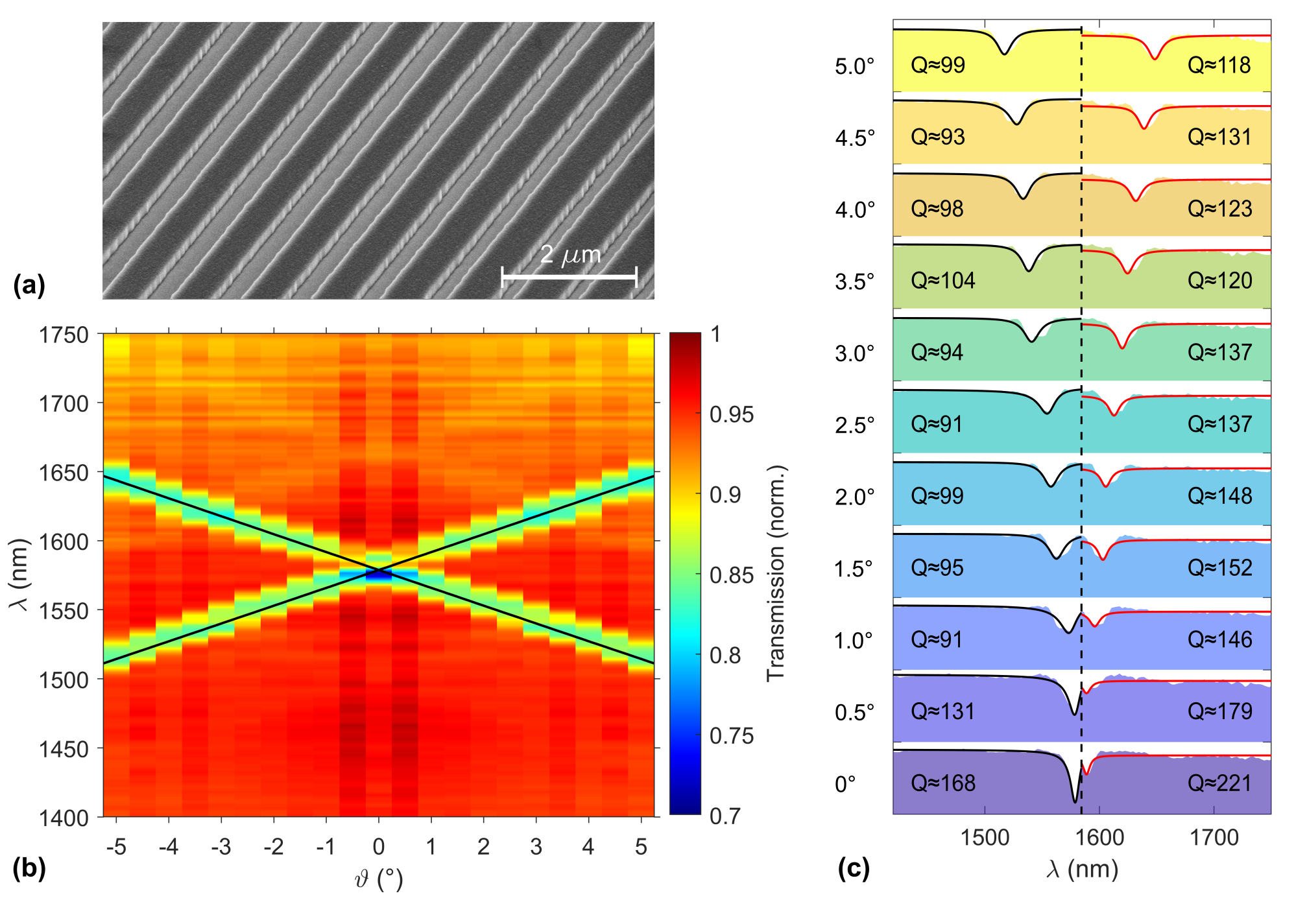}
\caption{(a) Scanning electron micrograph of the fabricated meta-grating. (b) Measured transmission spectrum of the metagrating in the telecom wavelength range for different incidence angles. Guided-mode resonances (GMR) are clearly appearing as minima in the transmission spectra, the bandgap is located at $\lambda_{\mathrm{gap}}$=1581~nm. The dispersion of the GMR closely follows the prediction based on the analytical model (black lines). Note that the measurements were carried out for incidence angles $\SI{0}{\degree}\leq\vartheta\leq\SI{5}{\degree}$. For the sake of better illustrating the band-structure, they are plotted mirror-symmetric, also including negative incidence angles. (c) Fits with Fano lineshapes to the measured linear spectra for varying incidence angles. Shaded areas are the measurement, where the baseline for all plots is set at $\SI{65}{\percent}$ transmission. Black solid lines are the Fano fit to the short-wavelength resonance, and red solid lines are the Fano fit to the long-wavelength resonance. The black dashed line marks the division between the fitting regions. The Q-factor calculated based on the Fano fit is indicated for each resonance.
\label{fig2}}
\end{figure*}

\subsection{Experimental Characterization of Co- and Counter-Propagating Pair-Generation}
    
\begin{figure*}[ht!]
\includegraphics[width=\textwidth]{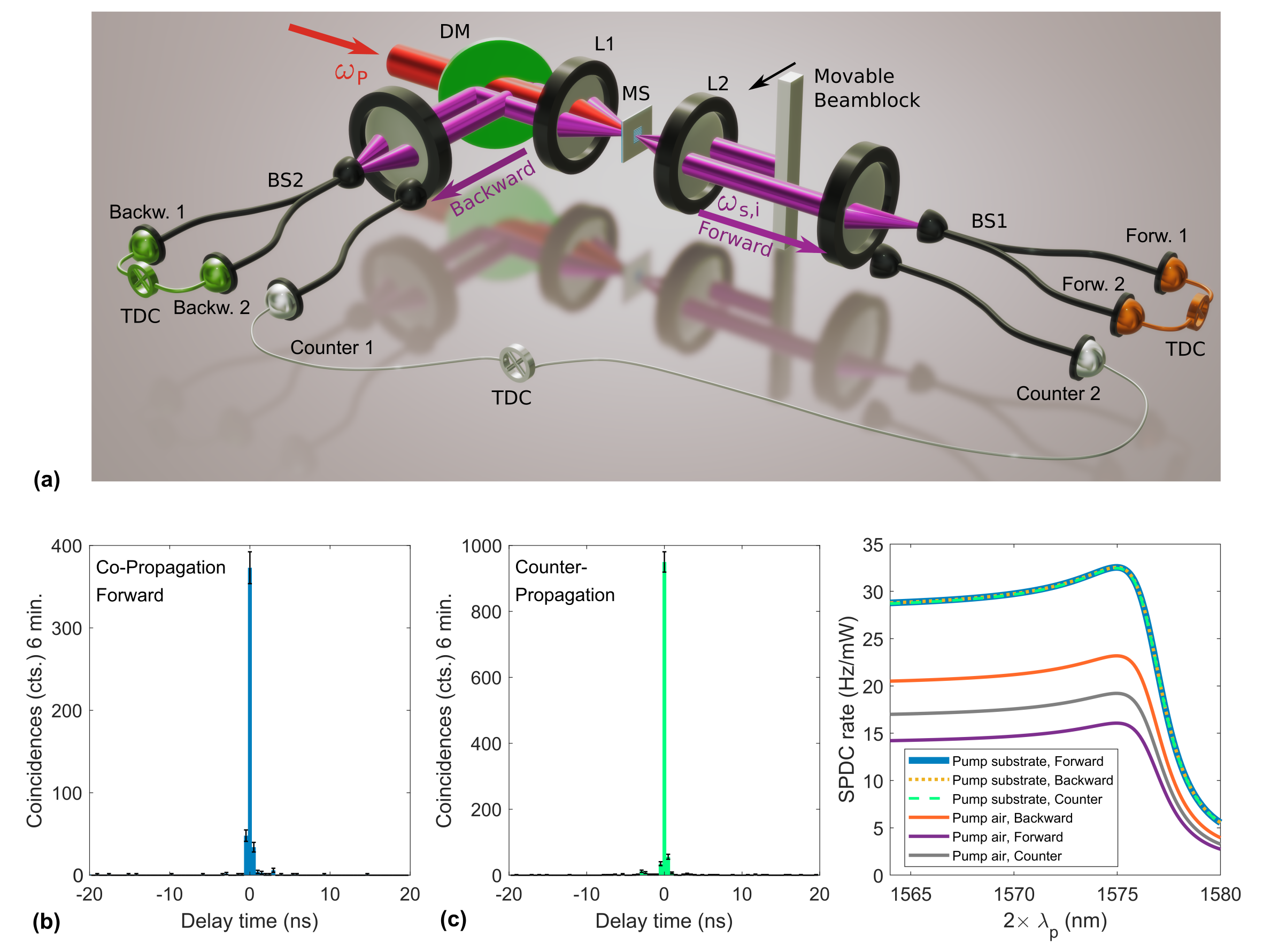}
\caption{(a) Schematic of the Hanbury Brown-Twiss interferometer used for coincidence measurements. A continuous-wave pump laser tunable in the range 782.7~nm~$\leq\lambda_p\leq$~789.5~nm is focused onto the metasurface by lens L1 (f=100~mm). Photon pairs in the forward direction are collected by lens L2 (f=50~mm), while pump photons are removed using a combination of interference filters (not shown). SPDC photons propagating in the backward direction are separated from the pump photons using a dichroic mirror (DM). For co-propagating collection geometries in the forward direction (detector position Forw. 1\&2, orange) or backward direction (detector position Backw. 1\&2, green), non-polarizing multimode fiber-beamsplitters are used. For counter-propagating detection geometry (detector position counter 1 \& 2, white), the photons are automatically separated. Temporal delays are measured using a time-to-digital converter (TDC). A movable beam block positioned in the back-focal plane of L2 allows the characterization of the propagation direction of photons in the transmission arm. (b)-(c) Coincidence histograms for (b) co-propagating forward and (c) counter-propagating detection geometry and excitation incident from the substrate side. The excitation power was P=91~mW at $\lambda_p=785$~nm and the exposure time $t=6$~min. Error bars mark the statistical uncertainty. (d) Simulated SPDC rate for varying pump wavelengths, excitation and detection geometries.
\label{fig3}}
\end{figure*}

We fabricate the designed nonlinear metagrating by deposition of SiO$_2$ on a commercial LN thin-film with quartz substrate~\cite{Fedotova:2022-3745:ACSP} and subsequent patterning of a $\SI{400}{\mu m}\times\SI{400}{\mu m}$ sized grating structure by etching of the SiO$_2$ layer only~\cite{zhangSpatiallyEntangledPhoton2022}. Since the LN waveguiding layer does not need to be patterned, scattering losses in the LN are kept at a minimal level. Figure~\ref{fig2}(a) shows a scanning electron micrograph of the fabricated structure. The linear properties of the metagrating are probed by transmission spectroscopy. For broadband illumination with a thermal lamp, pronounced dips in the transmission spectrum are visible, which spectrally shift with varying incidence angles $\vartheta$. The upper and lower branches of the modes, as well as the bandgap, are clearly visible. Overlaying the transmission plot with the analytically calculated mode dispersion confirms a very good agreement between the theoretically expected and experimentally obtained linear device properties. Note that for the calculation we use refractive index data measured by ellipsometry directly on our sample.

We further fit Fano line shapes to the experimentally measured transmission spectra, see solid lines in Fig.~\ref{fig2}(c), and calculate the Q-factor of the resonances in the fabricated device. The experimentally measured Q-factor for the short-wavelength resonance close to the band-gap is $Q\approx168$ and, at off-normal emission angles, slightly reduces to $Q\approx90\sim100$. The observed stability of the Q-factor over an extended angular range is a desirable property for directional tuning since it ensures efficient photon pair generation under various angles. Note that the modulation depth and Q-factor of the resonances observed in the transmission measurement are in part limited by the resolution of the available spectrometer (Ocean Insight NIRquest512 with full width at half maximum resolution of $\SI{3.1}{nm}$). Based on our simulations, we expect that the actual Q-factor is in fact above the values measured in this experiment, which is in-line with results from previous work \cite{zhangSpatiallyEntangledPhoton2022}. 

For the characterization of generated photon pairs, we use a Hanbury Brown-Twiss setup, see Fig.~\ref{fig3}(a), where pair-generation is detected by temporally coincident excitation of two multimode fiber-coupled single-photon avalanche diodes (SPADs). As a pump source, a tunable continuous-wave diode laser is used. In order to measure pair-generation in co- or counter-propagating emission configurations, correlations between the signals from the detectors in different positions are measured as marked in Fig.~\ref{fig3}(a). For pump photon suppression, we use a combination of interference long-pass filter (cut-on wavelength $\SI{1150}{nm}$) and a band-pass filter (central wavelength $\SI{1575}{nm}$, bandwidth $\SI{50}{nm}$) in forward and backward path. In sections S2 and S3 of the supplementary material, we provide more details on the experimental setup as well as a characterization of the pump laser.

We perform correlation measurements in co-propagation (forward) and counter-propagation geometry by exciting the metasurface with a pump beam of $\lambda_p=\SI{785}{nm}$ and power $P=\SI{91}{mW}$ incident from the substrate-side, compare the coincidence histograms in Figs.~\ref{fig3}(b),(c) (error bars mark the statistical uncertainty). We find for the emission of both photons in co-propagation and remarkably also for counter-propagating pair emission a high coincidence rate, marked by sharp correlation peaks around the zero-time delay bin. The coincidence rates $R_c$ extracted from the histograms are $R_{c,\mathrm{co}}=\SI{1.28}{Hz}\pm\SI{0.06}{Hz}$ for co-propagating emission in forward geometry and $R_{c,\mathrm{counter}}=\SI{2.92}{Hz}\pm\SI{0.09}{Hz}$ in counter-propagating geometry, respectively. The metagrating also generates a sizeable amount of photon pairs in the co-propagating backward direction, an exemplary histogram is found in supplementary material section~S4.
\newpage
Due to the technical limitations of our current experimental setup, we cannot characterize the angular properties of co-propagating photon pairs in the backward direction, and we don't consider this configuration in the following.

Of interest is now the ratio of the true photon pair rates $R_{c,\mathrm{true}}$ emitted in the different configurations. This can be estimated from the experimentally observed rate by accounting for the collection and detection efficiencies in both detection paths $\eta_{\mathrm{forw/backw}}$. However, in the case of a sub-wavelength thickness photon pair source, the absolute calibration of these efficiencies proves to be difficult. The presence of noise, e.g. from photoluminescence in LN \cite{sidorovPhotoluminescenceNominallyPure2020, okothMicroscaleGenerationEntangled2019}, inhibits the estimation of the collection efficiency via a simple calculation of the heralding efficiency \cite{sultanovTemporallyDistilledHighDimensional2023}. Therefore, we measure the transmission and collection efficiencies of a classical Gaussian beam at $\SI{1575}{nm}$, leading to estimated single photon collection efficiencies of $\eta_{\mathrm{forw}}\approx\SI{47}{\percent}$ in the forward path and $\eta_{\mathrm{backw}}\approx\SI{20}{\percent}$ in the backward path. The difference between the forward and backward arm is due to the different optical components in both beam paths, see supplementary material section~S2 for details.

Note that the coupling efficiency measured for a Gaussian beam will generally be considerably higher than for photon pairs generated by the nonlinear metasurface. Depending on the excited guided-mode resonance, pairs are emitted from the metasurface in two lobes with varying angular direction. Therefore, their overlap and correspondingly coupling efficiency to the fiber modes will be reduced as compared to a Gaussian beam. Furthermore, the pair coupling efficiency will vary with emission direction. For these reasons, the collection efficiency of $\eta_{\mathrm{forw}}\approx\SI{47}{\percent}$ should be regarded as an upper bound for the used experimental setup and will likely be considerably lower for photon pairs emitted by the metasurface. 
When accounting also for the quantum efficiency of the single-photon detectors of $\SI{25}{\percent}$ and the non-deterministic beamsplitter used only in the co-propagating configuration, we estimate the normalized pair-rates for the histograms shown in Figs.~\ref{fig3}(b,c) to be in the range of $\approx\SI{2}{Hz/mW}$ for the co-propagating forward geometry and about $\SI{5}{Hz/mW}$ for counter-propagating emission.\newpage
Since this value is based on the Gaussian beam calibration, it has to be considered a conservative estimate for the photon pairs from the metasurface, which is subject to a large error margin.
We also numerically calculate the generation efficiencies based on quantum-classical correspondence. In Fig.~\ref{fig3}(d), we show the simulated generation efficiency of degenerate SPDC for different pump wavelengths, collection geometries, and excitation scenarios, i.e. pump incident from air or substrate-side. Qualitatively, the pair-rate evolves with frequency the same for all excitation and detection scenarios. Away from the bandgap centered at $\lambda_{\mathrm{gap}}=\SI{1581}{nm}$ the generation efficiency is only weakly dependent on the pump wavelength and then significantly drops in the bandgap, where the density of optical states is reduced. An important factor is the incident direction of the pump field. For excitation from the air-side, the generation efficiency is reduced as compared to substrate-side excitation. In qualitative accordance with this, we obtain lower pair-generation rates in the experiment for air-side excitation of the metagrating than for the substrate-side excitation, see supplementary material.
\newpage
The generation rate of co-propagating photons in the backward direction is on a similar level as for the forward geometry, see supplementary material section~S4. Generally, the SPDC rate calculated in the simulations is about $6\times$ to $10\times$ larger than the SPDC rate estimated based on the experiment. We assign this largely to the difficulty in absolutely calibrating the collection and detection efficiency. Additionally, also unavoidable fabrication imperfections in the real sample may slightly decrease the device efficiency.

It needs to be highlighted that the CAR of our lithium niobate meta-grating source based on GMRs exceeds the CAR for many previously reported subwavelength thickness crystals and metasurfaces \cite{santiago-cruzPhotonPairsResonant2021,santiago-cruzResonantMetasurfacesGenerating2022,sonPhotonPairsBidirectionally2023,guoUltrathinQuantumLight2023,weissflogTunableTransitionMetal2023,saerensBackgroundFreeNearInfraredBiphoton2023} or microscale, non-phasematched devices \cite{okothMicroscaleGenerationEntangled2019,duongSpontaneousParametricDownconversion2022a} by one to two orders of magnitude - in particular given the operation at the relatively high pump power of $\SI{91}{mW}$. This is in line with previous results in GMR based metasurfaces \cite{zhangSpatiallyEntangledPhoton2022} and can be partly assigned to the wider band-gap of LN as compared to, e.g. (aluminum) gallium arsenide or gallium phosphide \cite{grinblatNonlinearDielectricNanoantennas2021}, which reduces photoluminescence for the here used pump wavelengths. Furthermore, the use of phase-matched pair-generation using high-Q guided-mode resonances in the metagrating is a very effective way to enhance SPDC while maintaining an ultra-thin design.

\subsection{Directional Tuning of Photon Pairs}
    
\begin{figure}
\includegraphics[width=0.85\columnwidth]{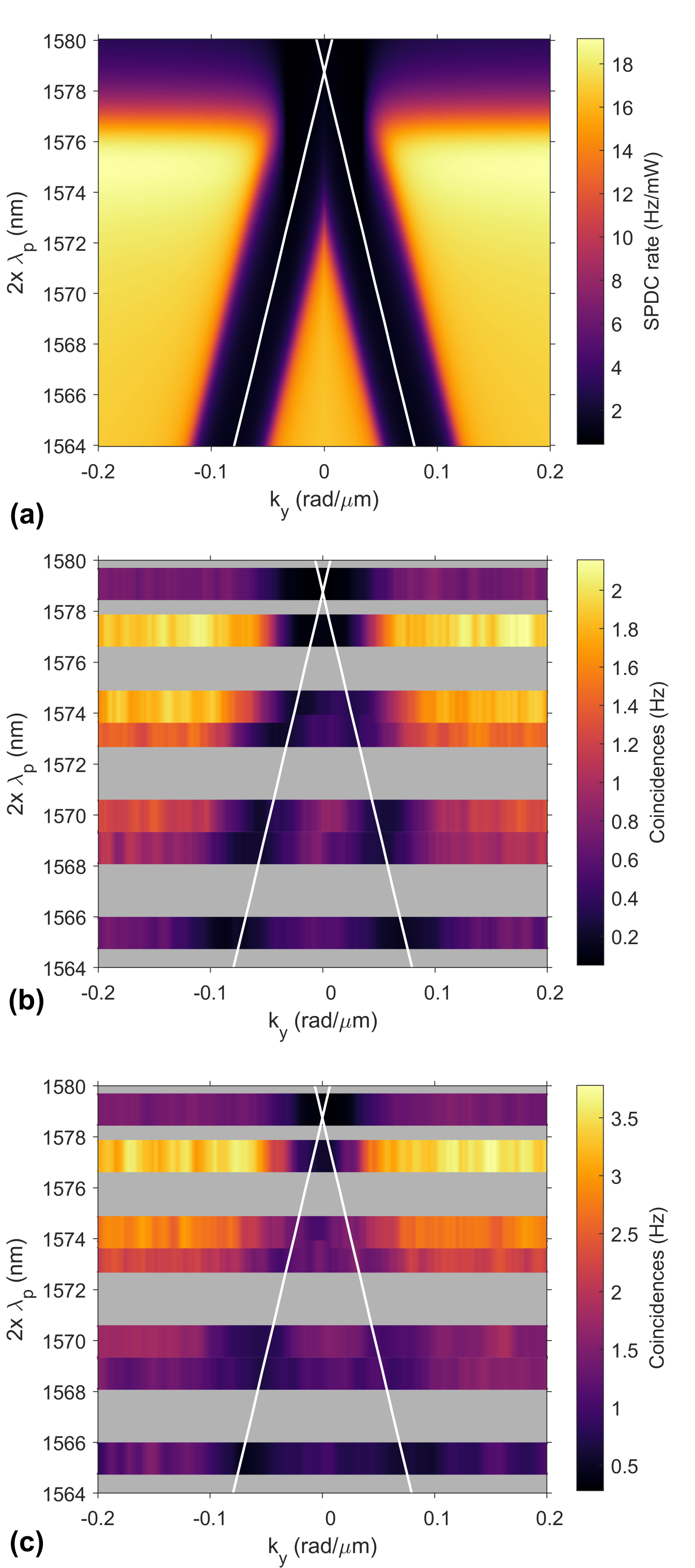}
\caption{(a) Simulation of SPDC rate in co-propagating forward direction depending on the transverse position of a beam block for different degenerate signal and idler wavelengths $\lambda_s=\lambda_i=2\times\lambda_p$. In this configuration, the beam block will prevent pair-detection whenever it is placed in the emission direction of the signal or idler photon, such that the photon pair propagation directions are marked by the minima of the detected pair rate. White lines mark the dispersion of the guided-mode resonances as calculated from the analytical model. (b) Experimentally measured coincidence rate in co-propagating forward geometry for different excitation wavelengths and beam-block positions. The emission direction coincides with the transverse momentum of the guided-mode resonances (white lines). Grey bars indicate wavelength regions that are inaccessible due to the mode-hopping of the pump lasers. (c) Experimentally measured SPDC rate in counter-propagating geometry for different beam-block positions. The emission angle again closely follows the position of the guided-mode resonances (white lines).
\label{fig4}}
\end{figure}

In the next step, we demonstrate the ability to tune the emission angle of photon pairs. Since the flux of the photon pairs is too low for direct back-focal plane imaging with a classical camera, we instead use a movable beam block with size $\SI{700}{\mu m}$ placed in the back-focal plane of the collimation lens in forward direction (lens L2, see Fig.~\ref{fig3}(a)). By moving the beam block across the emission lobe of first the signal and then the idler photon, the detected coincidence rate will drop whenever the beam block is partly or fully obscuring the respective beam-path. Therefore, a decrease in coincidences marks the emission angle of the photon pairs. Since the beam block is placed in the back focal plane, the lateral position can be directly converted to the transverse momentum of the photon pairs. We use a pump beam incident from the substrate-side, which provides the highest pair-generation efficiency based on our previous analysis. The expected dependence of the pair-generation rate on the beam-block position is shown in Fig.~\ref{fig4}(a), where the emission angle of the photons increases for a larger detuning from the bandgap at $\SI{1581}{nm}$. The emission direction is identified by clear minima in the pair rate, which appear at angles dictated by the dispersion of the guided modes (see white lines in Fig.~\ref{fig4}(a)). For the experiment, we use a diode laser tunable in the wavelength range from $\SI{782.7}{nm}\leq\lambda_p\leq\SI{789.5}{nm}$. However, the tunability of the laser is limited by mode hopping, which reduces the number of available pump wavelengths.
\newpage
Wavelength regions that are inaccessible due to mode hopping are marked with grey bars in the experimental plots Figs.~\ref{fig4}(b,c). A detailed characterization of the laser is provided in supplementary material section S3.

We measure coincidence histograms for many pump wavelengths and beam-block positions and extract the coincidence counts. In Figs.~\ref{fig4}(b,c), we plot the evolution of the detected coincidence rate with the beam block position for different excitation wavelengths. Note that we use a moving average filter to partly compensate for fluctuations in the coincidence rate due to the photon statistics. Refer to section S5 of the supplementary material for the raw data.

The tuning of the emission angle with pump wavelength is clearly visible. For the co-propagating detection configuration in Fig.~\ref{fig4}(b), two minima in the coincidence rate at opposite transverse wave vectors $\pm k_y$ mark the directions where the beam block crosses the emission path of the signal or idler photon. In excellent agreement with the simulation, these emission directions shift to larger angles for decreased pump wavelength. Furthermore, the emission direction of the photon pairs follows the dispersion relation of the GMR frequencies (compare white lines in Figs.~\ref{fig4}(b,c)). We note a reduction of the detected pair-rate for larger emission angles, which is somewhat stronger than expected from the simulation. This is mainly due to the reduced fiber-coupling efficiency of the photon pairs in the experiment, which degrades with increased propagation angle.

With these results, we demonstrate that the design of our metagrating is very effective in controlling the spatial properties of the generated photon pairs. Note that in the case of degenerate, co-polarized and co-propagating SPDC, the definition of signal and idler mode is done based on the spatial propagation direction. The signal photon mode is, e.g. defined to have a wave-vector $-\left|k_y\right|=k_s$ and the idler photon correspondingly $+\left|k_y\right|=k_i$ in the same half-space. When turning to a counter-propagating detection geometry, both photons are emitted into opposite half-spaces, which can be used to distinguish signal and idler photons. Now, when carrying out the same emission angle analysis for counter-propagation, see Fig.~\ref{fig4}(c), we interestingly find two minima again when scanning the beam block. This shows that the single photon emitted into the signal half-space is in a superposition state of propagating along $k_s=\pm\left|k_y\right|$. Due to momentum conservation, the idler photon propagates into opposite direction $k_i=\mp\left|k_y\right|$, corresponding in total to a path-entangled state $\ket{\psi}=1/\sqrt{2}\left(\ket{\left|k_y\right|,-\left|k_y\right|}+e^{i\alpha}\ket{-\left|k_y\right|,\left|k_y\right|}\right)$. Due to the grating symmetry with respect of $y \rightarrow -y$, we theoretically expect that $\alpha = 0$ at the degenerate wavelengths of signal and idler photons, although with our experimental configuration we cannot measure $\alpha$ which would require tomographically complete set of measurements on all paths \cite{shadboltGeneratingManipulatingMeasuring2012}. There is an interesting potential to tune $\alpha$ in the non-degenerate regime and by varying the unit cell symmetry.
We envision this to be a promising platform for further investigations, e.g. for the generation of hyper-entangled states by also including the frequency degree of freedom.

\section{Conclusions}

In this work, we designed a nonlinear lithium niobate metagrating and experimentally demonstrated photon pair generation with angular tunability in co- as well as counter-propagating emission. To the best of our knowledge, this is the first demonstration of such precise control over the spatial properties of nonlinearly generated photon pairs from a metasurface. We show that the emission angle of co- and remarkably also counter-propagating photon pairs can be finely tuned by selectively generating signal and idler in guided-mode resonances with high angular dispersion. Moreover, we present a straightforward design strategy for flat, nonlinear SPDC sources that generate tunable and counter-propagating pairs. We, in particular, highlight the latter since generating pairs counter-propagating with respect to the pump direction in a phase-matched nonlinear device is usually a challenging task that requires very careful modal engineering \cite{saraviGenerationCounterpropagatingPathEntangled2017} or quasi-phasematching schemes with very small poling periods \cite{kuoPhotonpairProductionFrequency2023}. Here we have shown that by using transversal phase-matching of guided-modes and probabilistic out-coupling into two opposing half-spaces, counter-propagating pairs can be generated with a very straightforward device design. It is important to note that in our nonlinear metagrating for all detection configurations always, both signal and idler photons are generated on resonance. This is very favorable for the generation efficiency and contrasts with a recent demonstration of counter-propagating pair-emission from a BIC-resonant metasurface \cite{sonPhotonPairsBidirectionally2023}, where one photon of the pair is generated off-resonantly to achieve counter-propagation.

In this work, we control the emission angle all-optically by means of tuning the excitation wavelength. However, since the propagation constant of the guided modes and, therefore, the emission angle sensitively depends on the refractive index of the guiding layer, using the Pockels effect in LN could also be leveraged to electrically modulate the emission direction while keeping the pump wavelength constant. In our experiment, the angular tuning range was only limited by the tunability of the pump laser and the coupling efficiency to the fiber-based detectors. In principle, the underlying physics of guided-mode resonances in our metagrating allows angular tuning in much larger ranges, as already indicated by the linear characterization of the device. Furthermore, the CAR of about 7500 in counter-propagating geometry is significantly higher than for many other sub-wavelength thickness crystals or metasurfaces reported so far, which is an important step towards higher practicability of metasurface-based photon pair sources.
\newpage
Nonetheless, for most use cases, the generation efficiency needs to be further increased. This can, for instance, be realized by also resonantly coupling the pump wave to the wave-guiding layer and employing higher-order quasi-phasematching using periodic poling of the nonlinear metasurface \cite{fedotovaSpatiallyEngineeredNonlinearity2023} or by optimizing the Q-factor of the guided modes \cite{jiangTunableSecondHarmonic}.

In summary, with this work, we demonstrate precise spatial control of photon pairs generated in a nonlinear metasurface, which adds an important building block to the toolset of subwavelength thickness quantum light sources.


\section*{Funding}
This work was supported by the Australian Research Council (\url{https://doi.org/10.13039/501100000923}) Centre of Excellence for Transformative Meta-Optical Systems (CE200100010) and the Australia–Germany Joint Research Cooperation Scheme of Universities Australia (UA) and the German Academic Exchange Service (DAAD \url{https://doi.org/10.13039/501100001655}, Grant No. 57559284). The authors acknowledge financial support from the Deutsche Forschungsgemeinschaft (DFG, German Research Foundation, \url{https://doi.org/10.13039/501100001659}) through the International Research Training Group Meta-ACTIVE (IRTG 2675, project number 437527638) and the Collaborative Research Center NOA (CRC 1375, project number 398816777) and project Megaphone (project number 505897284). This project was funded by the European Union’s Horizon 2020 Research and Innovation program (\url{https://doi.org/10.13039/501100007601}) METAFAST (Grant Agreement No. 899673) and by the German Federal Ministry of Education and Research BMBF (\url{https://doi.org/10.13039/501100002347}) through project QOMPLEX (13N15985) and project PhoQuant (13N16108). S.S. acknowledges funding by the Nexus program of the Carl-Zeiss-Stiftung (\url{https://doi.org/10.13039/100007569}) (project MetaNN). This work used the ACT node of the NCRIS-enabled Australian National Fabrication Facility (\url{https://doi.org/10.13039/100008015}, ANFF-ACT).

\section*{Author Contributions}
All authors have accepted responsibility for the entire content of this manuscript and approved its submission. M.A.W., J.M., J.Z., F.S, and A.A.S. conceived the initial idea. M.A.W., J.Z. and T.F. performed analytical and numerical calculations. J.Z. fabricated the sample and performed linear transmission measurements. M.A.W. and J.M. built the HBT-setup for correlation measurements. M.A.W. performed the correlation measurements, analyzed the experimental data, created all figures and wrote the first draft of the manuscript. All authors discussed the results and revised the manuscript. A.A.S., F.S., T.P., D.N.N. and S.S. provided overall supervision of the project and acquired funding. 

\section*{Conflict of Interest}
Authors state no conflict of interest.

\section*{Data Availability}
The datasets generated during and/or analyzed during the current study are available from the corresponding author on reasonable request.

\bibliography{bibliography}

\end{document}